\documentstyle[12pt,version2,aps]{revtex}
\begin{document}
\renewcommand{\theequation}{\arabic{section}.\arabic{equation}}
\newcommand{\eqreset}{\setcounter{equation}{0}}
\begin{center}
{\large\bf The Free Energy and the Scaling Function \\
of
the Ferromagnetic Heisenberg Chain \\
in a Magnetic Field\footnote{
  {\normalsize Submitted to J.Phys.Soc.Jpn}
}
 } \vspace{5mm}\\
 Hiroaki {\sc Nakamura}\footnote{e-mail "hiroaki@eagle.issp.u-tokyo.ac.jp"}
         and
    Minoru {\sc Takahashi}
 \vspace{5 mm}\\
        {\it Institute for Solid State Physics,University of Tokyo}\\
        {\it Roppongi,Minato-ku,Tokyo 106} \\
(received ~~~~~~~~~~~~~~~~~~  )
\end{center}
\vspace{5 mm}
\begin{abstract}
\newcommand{\qbSpin}{\mbox{\boldmath $S$}}
\newcommand{\qbspin}{\mbox{\boldmath $s$}}
\newcommand{\bSpin}{\mbox{\boldmath $R$}}
\newcommand{\bspin}{\mbox{\boldmath $r$}}
A nonlinear susceptibilities (the third derivative
of a magnetization $m_S$ by a magnetic field $h$ ) of the $S$=1/2 ferromagnetic
Heisenberg chain and the classical Heisenberg chain
are calculated at low temperatures $T.$
In both chains the nonlinear susceptibilities diverge as
$T^{-6}$
and a linear susceptibilities diverge as $T^{-2}.$
The arbitrary spin $S$ Heisenberg ferromagnet $[$
${\cal H} = \sum_{i=1}^{N} \{  - J \qbSpin_{i} \qbSpin_{i+1}
             -  (h/S)  S_{i}^{z}  \}$ $(J>0),$ $]$
has a scaling relation
between
$m_S,$ $h$ and $T:$
$m_S(T,h) = F( S^2 Jh/T^2).$
The scaling function
$F(x)$=(2$x$/3)-(44$x^{3}$/135) + O($x^{5}$)
is common to all values of spin $S.$

\vspace{10mm}
\end{abstract}
\pagebreak
\section{Introduction}\label{intro}
In the 1970's integral equations have been proposed
to calculate thermodynamic quantities of the $S$=1/2 Heisenberg
chain\cite{TMstr1,Gaudin}.
In this calculation the Bethe Ansatz method
was used to diagonalize the Heisenberg Hamiltonian.
One of the authors has calculated the Gibbs free energy
numerically by the integral equations\cite{TY}.

On the other hand,
the other approach has been developed.
In this approach a quantum one-dimensional system is transformed
to a classical two-dimensional system by the Trotter formula.
Betsuyaku has defined a transfer matrix
and he has calculated an energy and a specific heat numerically
of the $S$=1/2 XY model and the $S$=1/2 Heisenberg model\cite{betsu}.
By diagonalizing the transfer matrix with the Bethe Ansatz method,
Koma\cite{Koma} has calculated
the Gibbs free energy and a magnetic susceptibility
of the $S$=1/2 Heisenberg chain.
These quantities coincide numerically with results of ref.3.
Moreover a correlation length of the $S$=1/2 Heisenberg chain
has been calculated numerically\cite{Yamada}.
Furthermore Suzuki,Akutsu and Wadati\cite{SAW} pointed out
that the transfer matrix of
the $S$=1/2 XYZ model was a special case
of the inhomogeneous eight-vertex model
which was solved by Baxter\cite{Baxter}.
One of the authors\cite{TMxyz} has obtained
the Gibbs free energy and the correlation length
of the $S$=1/2 XXZ model in a magnetic field
by solving self-consistent equations for infinite numbers,
which were derived by the Bethe Ansatz method.
\newcommand{\qbSpin}{\mbox{\boldmath $S$}}
\newcommand{\qbspin}{\mbox{\boldmath $s$}}
\newcommand{\bSpin}{\mbox{\boldmath $R$}}
\newcommand{\bspin}{\mbox{\boldmath $r$}}

In this paper we try to calculate the free energy
of the $S$=1/2 ferromagnetic Heisenberg chain in a magnetic field
by solving self-consistent equations for infinite numbers.
We also calculate the linear and the third order susceptibilities
of the classical Heisenberg chain analytically.
By comparing these results,
we propose a scaling function
of the arbitrary spin $S$ Heisenberg chain
defined by
\begin{equation}
{\cal H} = - J \sum_{i=1}^{N}
                \qbSpin_i \qbSpin_{i+1}
             -   \frac{ h}{S} \sum_{i=1}^{N} S_{i}^{z} ,  \label{eq1}
\end{equation}
where $\qbSpin_{i} = (S_{i}^{x},S_{i}^{y},S_{i}^{z})$
are spin operators
and $J$ is the positive coupling constant.
A partition function $Z_{N,S},$
the Gibbs free energy per site $f_S$
and a magnetization $m_S$ of this model are
defined as follows:
\begin{eqnarray}
Z_{N,S} &\equiv&  {\rm Tr} \left\{
              \exp \left(- \frac{ {\cal H} }{ T } \right)
                  \right\}  ,  \label{partZ} \\
f_S &\equiv& - \lim_{ N \rightarrow \infty }
               \frac{T}{N} \ln Z_{N,S}      ,\label{1free}  \\
m_S &\equiv&   -\frac{\partial f_S}{\partial h}
.\label{magd}
\end{eqnarray}
A linear susceptibility $\chi_{1,S}$
and the third order susceptibility $\chi_{3,S}$
are defined as follows:
\begin{eqnarray}
\chi_{1,S}
    &\equiv& \left. -\frac{{\partial}^{2}f_S}{\partial h^{2}} \right|_{h=0}
,\label{defchi1.1} \\
\chi_{3,S}
    &\equiv& \left. -\frac{1}{3!}\frac{{\partial}^{4}f_S}{\partial h^{4}}
\right|_{h=0} \  . \label{defchi3.1}
\end{eqnarray}

In \S2
we calculate a scaling function of the
magnetization in the classical limit.
Moreover we extend this scaling function to the case of an
arbitrary spin.
In \S3
to check our extended scaling function for $S$=1/2,
we calculate the scaling function numerically
by the Bethe Ansatz method.
We calculate the Helmholtz free energy and the
third order nonlinear susceptibility.
We compare numerical results with the extended scaling function.
In \S4
we summarize the scaling function of the spin $S$
ferromagnetic Heisenberg chain and
 new results of the numerical calculation for $S$=1/2.
In Appendix we calculate the third order nonlinear susceptibility
for the classical Heisenberg chain analytically.

We define the classical Heisenberg Hamiltonian ${\cal H }^{\rm C}$
with classical spins $\bspin_{i}$
as follows:
\begin{equation}
	{\cal H }^{\rm C} =
 -J_{0} \sum_{i=1}^{N} {\bspin}_{i}\cdot {\bspin}_{i+1}
             -h \sum_{i=1}^{N} {r}_{i}^{z}.  \label{1.1}
\end{equation}
In the classical chain,
a partition function $Z_{N}^{\rm C} ,$
the Gibbs free energy per site $f^{\rm C},$
a magnetization $m^{\rm C}$
and a susceptibilities
$\chi_{1}^{\rm C},\chi_{3}^{\rm C}$
are defined
by changing from ${\cal H}$ to ${\cal H^{\rm C} }$
in eqs.(\ref{partZ})-(\ref{defchi3.1}).
\eqreset
\section{The Scaling Function at an Arbitrary Spin}\label{conj}
\subsection{Analytical Calculations in the Classical Limit ($S \rightarrow
\infty$)}\label{con-1}

Let's consider the classical limit (i.e.$S \rightarrow \infty$)
of the ferromagnetic Heisenberg chain in a magnetic field.
We define normalized spin operators $\qbspin_i$
as follows:
\begin{equation}
\qbspin_i \equiv \frac{\qbSpin_i}{S}. \label{sspin}
\end{equation}
By using $ \qbspin_i $'s,
the Hamiltonian defined by eq.(\ref{eq1})
is written as follows:
\begin{equation}
{\cal H} = - J_0 \sum_{i=1}^{N}
                \qbspin_i \qbspin_{i+1}
             -   h \sum_{i=1}^{N} s_{i}^{z} ,  \label{eq11}
\end{equation}
where
\begin{equation}
J_0 = JS^2 . \label{3.4.1}
\end{equation}
We can regard the spin operators $\qbspin_i$
as classical three dimensional unit vectors $\bspin_i$
in the classical limit,
because the spin operators $\qbspin_i$ become commutable.
( For example, $[s_i^x,s_i^y] = i \hbar s_i^z/S \rightarrow 0$
in the limit of $S \rightarrow \infty$
on condition that $JS^2$ is fixed. )
Therefore,
to calculate
susceptibilities in the classical limit\cite{Fisher},
we can use the classical Hamiltonian ${\cal H }^{\rm C}$
defined by eq.(\ref{1.1})
instead of the quantum Hamiltonian of eq.(\ref{eq11}).
As $T/J_0 \rightarrow 0,$
we get susceptibilities as follows:
\begin{equation}
   \chi_{1,S=\infty}  \rightarrow \frac{2J_{0} }{3T^{2} } ,\label{1linearchi}
\end{equation}
\begin{equation}
\chi_{3,S=\infty}
   \rightarrow - \frac{44J_{0}^{3}}{135T^{6}} . \label{1chiinfty}
\end{equation}
They are calculated from the 2-point and 4-point correlation functions at
$h$=0.
Details are given in Appendix.
{}From eqs.(\ref{1linearchi}) and (\ref{1chiinfty}),
we get a scaling relation between the magnetization,
the magnetic field and the temperature
as follows:
\begin{equation}
 m_{S=\infty} (T,h)
 =  F\left( \frac{  J_{0} h }{T^{2}} \right) ,\label{3.4.0.1}
\end{equation}
where
\begin{equation}
F(x) =  \frac{2}{3} x - \frac{44}{135} x^{3}
                     + O (x^{5}) . \label{4.3.3}
\end{equation}

\subsection{The Scaling Function at an Arbitrary
Spin from the Classical Limit }\label{con-2}

In the previous subsection
we get the scaling function in the infinite spin limit
where we can use the classical Hamiltonian
to calculate the scaling function.
In this subsection
we consider the finite spin case.
At low temperatures,
the correlation length becomes large
and the system is nearly ordered.
Excitations behaves like spin waves
independently of $S.$
This is the reason why the finite $S$
Heisenberg chain belongs
to the same universality class
as the infinite spin chain
(i.e. the classical chain).
Therefore
the classical Heisenberg chain becomes a good model
to calculate the scaling function of the magnetization for
the spin $S$ quantum chain.
On the basis of this physical picture,
we expect the scaling relation
of the arbitrary spin $S$ Heisenberg chain
at sufficient low temperatures and $(S^2 Jh/T^2 \ll 1)$
as follows:
\begin{equation}
  m_S(T,h)
 =  F\left( \frac{  S^2J h }{T^{2}} \right) ,\label{arbS}
\end{equation}
where the scaling function $F(x)$ is eq.(\ref{4.3.3})
 for the arbitrary spin $S.$
In the next section we check the extended scaling function
for $S$ =1/2 by comparing with the numerical result given by the
Bethe Ansatz method.
\eqreset
\section{The Scaling Function at $S$ = 1/2 }\label{num}
\subsection{Methods of Numerical Calculations at $S$ = 1/2 }\label{num1}
We calculate numerically the Gibbs free energy and the magnetization
of the $S$=$1/2$ Heisenberg chain in a magnetic field.
One of the authors\cite{TMxyz}
has given the following equations
with infinite numbers $  p_l $'s
\begin{eqnarray}
& & p_{l}= \frac{4T}{J}
      \left[ \frac{h i }{T} + \pi (l- \frac{1}{2} )
             + \frac{1}{2i} X_{l}
      \right], \label{Lxy}
 \\
& &  X_{l} \equiv \left[  \frac{J p_{l} i}{2T( p_{l}^{2} +1 )} \right]
             + \sum_{j=1}^{ \infty }
            \ln L (p_{l},p_{j}) L(p_{l},- \bar{ p_{j} } ),  \label{pl} \\
& & l = 1,2,3,\cdots,\infty , \nonumber
\end{eqnarray}
where
\begin{equation}
  L(x,y) \equiv
           \frac{ iy+ \left[ 1- \frac{1}{ (1-ix)} \right] }{
                 -iy+ \left[ 1- \frac{1}{ (1+ix)} \right] }    . \label{Lxy2}
\end{equation}
These $ p_{l}$'s give the maximum eigenvalue
$ \Lambda^{0} ( T/J, h ) $ of the transfer matrix and
the Gibbs free energy per site as follows:
\begin{eqnarray}
\Lambda^{0} ( T/J, h ) &=& 2 \prod_{l=1}^{ \infty }
            {  [ \frac{ J   }{ 4 \pi T ( l- \frac{1}{2} ) } ]  }^{2}
            {  [  ( p_{l}^{2} + 1 ) ( { \bar{ p_{l} } }^{2} + 1 ) ] }^{1/2} ,
   \label{eigen} \\
f_{S=1/2}&\equiv& f(T/J,h)  = -T \ln \Lambda^{0} (T/J,h ) .  \label{free}
\end{eqnarray}

The magnetization is given by a differentiation of
the free energy with respect to $ h.  $
We need to calculate $ \partial p_{l}/\partial h. $
We define $q_{l}$'s as follows,
\begin{equation}
 q_{l} \equiv \frac{\partial p_{l} }{\partial h } ,
 \bar{ q_{l} } \equiv  \frac{\partial \bar{ p_{l} }  }
{\partial h }  . \label{q}
\end{equation}
Differentiating eqs.(\ref{Lxy})-(\ref{Lxy2}),
we obtain the following equation for $ q_l $'s:
\begin{equation}
q_{l} = \frac{4T}{J}
              \left[ \frac{i}{T}
                   + \frac{J(1- p_{l}^{2}) }
                     {4T { (p_{l}^{2} +1 )}^{2} }
                   + \frac{1}{2} Y_{l}
              \right] ,                    \label{dot}
\end{equation}
where
\begin{equation}
Y_{l}\equiv \sum_{j=1}^{\infty}
              \left[
                   \alpha(p_{l}, p_{j}          )      q_{j}
                 - \beta(p_{l}, p_{j}          )      q_{l}
                 - \alpha(p_{l},- \bar{ p_{j} } )\bar{ q_{j} }
                 - \beta(p_{l},- \bar{ p_{j} } )\bar{ q_{l} }
               \right],
                                     \nonumber
\end{equation}
\begin{eqnarray}
\alpha(x,y) &\equiv&   \frac{1}{ iy+(1-\frac{1}{1-ix} ) }
           + \frac{1}{-iy+(1-\frac{1}{1+ix} ) }, \label{m1} \\
\beta(x,y) &\equiv&   \frac{ \frac{1}{ {(1-ix)}^{2} } }
                  { iy+(1-\frac{1}{1-ix} ) }
           + \frac{ \frac{1}{ {(1+ix)}^{2} } }
                  {-iy+(1-\frac{1}{1+ix} ) }
              . \label{n1}
\end{eqnarray}
The magnetization per site is represented by $ p_{l}$'s and $ q_{l}$'s:
 \begin{eqnarray}
 m_{S=1/2} &\equiv& m
 =   2T \sum_{l=1}^{\infty }
              {\rm Re} \frac{ p_{l} q_{l} }{  p_{l}^{2} + 1 } . \label{mag}
\end{eqnarray}
In actual numerical calculations we can approximate
$ p_{l}$'s and $ q_{l}$'s as follows\cite{TMxyz}:
\begin{eqnarray}
  p_{l} &=& \frac{4 T \pi}{J}  (l - \frac{1}{2} ) + i  {\rm Im } (p_{L})  ,
   \label{plgeq}  \\
    q_{l} &=& q_{ L } , \label{qlgeq}
\end{eqnarray}
for $l > L,$ where $L$ is an integer.
We can obtain solutions of
$p_1 ,\cdots, p_L$ and $q_1,\cdots,q_L.$
By increasing $L,$
we can estimate the product in eq.(\ref{eigen}) and
the summation in eq.(\ref{mag}).
\subsection{Numerical Results at $S$ = 1/2 }\label{result}
\subsubsection{The Expansion of the Free Energy by the Least-Squares
Method}\label{result-1}

Let us consider the Helmholtz free energy $g(t,m)$ per site of the Heisenberg
chain in a magnetic field which is defined as follows:
\newcommand{\IT}{\left( \frac{1}{t} \right) }
\newcommand{\TTT}{\left( \frac{h}{Jt^2} \right) }
\newcommand{\xTTT}{\left( \frac{Jh}{T^2} \right) }
\begin{equation}
g(t,m) \equiv f(t,h ) + m h     ,     \label{gtm}
\end{equation}
where
\begin{equation}
t \equiv \frac{T}{J}. \label{ttt}
\end{equation}

In the case of $m=0,$
at low temperatures
it has been already obtained by several authors\cite{TY,TMMSW}.
Takahashi and Yamada\cite{TY} obtained
\begin{equation}
g(t,0) =Jt^{1.5}\left\{ -1.042 + 1.00t^{0.5} -0.9t+O(t^{1.5})
           \right\}
 .\label{TY1}
\end{equation}
\begin{equation}
\chi_{1,S=1/2} =
      \frac{1}{Jt^2} \left\{ 0.1667+ 0.581 t^{0.5} + 0.68t+O(t^{1.5}) \right\}
.\label{TY2}
\end{equation}
{}From the modified spin wave theory\cite{TMMSW}
the susceptibility equals:
\begin{equation}
\chi_{1,S=1/2}  =
      \frac{1}{Jt^2} \left\{ (1/6) + 0.582597 t^{0.5} + 0.678839 t
+O(t^{1.5})\right\}.
				\label{kai4}
\end{equation}

When $m$ is finite,
the coefficients of the free energy and the magnetization
will be changed from the values of the $m$=0 case.
By analyzing our numerical data,
we can get the  $m$ dependence of the free energy as follows:
\begin{equation}
\frac{g(t,m)}{J } =     a_{0} ( t  ) t^{1.5}
            +  t^{2} \{
                   a_{1} ( t  ) m^{2}
            +      a_{2} ( t ) m^{4}
            +      O( m^{6} )
                        \}    ,\label{g}
\end{equation}
where
\begin{eqnarray}
 a_{0} ( t ) &=& -1.04218 + 1.00 t^{0.5} -0.94 t +0.9t^{1.5},
                               \label{kai0} \\
 a_{1} ( t ) &=& \frac{1}{2} { \left\{
		\frac{1}{ (1/6) + 0.5826t^{0.5} + 0.678 t +O(t^{1.5}) }
			\right\} }  ,
				\label{kai1} \\
 a_{2} ( t ) &=& \frac{1}{4}  { \left\{
		\frac{1}{ 0.153 + 0.83 t^{0.5} + 2.4 t +O(t^{1.5}) }
			\right\} } .
				\label{kai2}
\end{eqnarray}
We show the $t^{0.5}  $ dependence of $ g (t,m)/(J{t}^{3/2})$ at $ m = 0,$
0.28, 0.632 in Fig.\ref{free_ene}.
This figure gives us
an important result
that the $m$ dependence of $g(t,m)$ does not begin with
$t^{1.5}$ but $t^2.$
We plot $a_0(t),$ $a_1(t)$ and $a_2(t)$
versus $t^{0.5}$ in Figs.\ref{keisu1}-\ref{keisu3},
where solid curves are given by the least-squares method
and where filled circles denote numerical data.

Since $h = \partial g/ \partial m,$
eq.(\ref{g}) gives us the relation between $m$ and $h$
at fixed values of $t$ as follows:
\begin{equation}
\frac{h(t,m)}{Jt^2} =  2  a_{1}(t) m
            +     4  a_{2}(t) m^{3}
            +           O( m^{5} )
	             .\label{h}
\end{equation}
When $h/(Jt^2) \ll 1,$
we can transform eq.(\ref{h}) into the following equation:
\begin{equation}
m = \frac{1}{2a_{1}(t)} \TTT - \frac{a_{2}(t)}{4a_1(t)^4} \TTT^3 + O\TTT^5.
\label{chi0}
\end{equation}
By differentiating eq.(\ref{chi0}) by $h,$
we have the linear susceptibility
as follows:
\begin{eqnarray}
\chi_{1,S=1/2} &=& \frac{1}{ 2 Jt^{2}  a_{1} (t) }  \nonumber \\
       &=& \frac{1}{Jt^{2} } \left\{
	               (1/6) + 0.5826t^{0.5} + 0.678 t +O(t^{1.5})
			\right\} . \label{chi1}
\end{eqnarray}
This result is consistent with eqs.(\ref{TY2}) and (\ref{kai4}).
Moreover we get the third order susceptibility as follows:
\begin{eqnarray}
\chi_{3,S=1/2}
	 &=& -  \frac{a_2 (t)}{4 J^{3} t^{6} a_{1} (t)^{4}  }
        \nonumber \\
       &=& - \frac{1}{J^{3} t^{6} } \left\{
	               0.00504 + 0.0431 t^{0.5} + 0.13 t +O(t^{1.5})
			\right\} .
        \nonumber \\
 & & \ \ \label{chi3}
\end{eqnarray}
\subsubsection{The Magnetization Curve } \label{result-2}
In eq.(\ref{chi0}),
we can get a relation
between $m$ and $h,$
at the given temperature $t.$
Here we consider how this relation
behaves in the limit of $t \rightarrow 0.$
Equation (\ref{mag}) gives us $m$ at
fixed values of bath the temperature and
the magnetic field.
In Fig.\ref{m_hcurve}, we plot $h/(Jt^{2})$ versus
$t^{0.5}$ for 8 values of $m:$ $m$ = 0.01,0.04,$\cdots,$ 0.63.
{}From this figure,
it is clear that $\ln (h/Jt^2 )$ scale with $t^{0.5}.$
The extrapolated values given by Fig.\ref{m_hcurve}
are shown by empty squares in Fig.\ref{m_hcurve2},
where a bold curve is drawn as a guide for eyes.
Moreover in this figure we show $m$ versus $\ln (h/Jt^2)$
for 5 values of $t:t$ = 0.2,$\cdots,$0.005
by filled circles.
This figure tells us that as $t\rightarrow 0,$
magnetization curves are going to a bold curve,
which is regarded as the scaling function.
Especially for small $h/(Jt^2),$
the scaling relation between the magnetization,
the magnetic field and the temperature is
the following equation:
\begin{equation}
  m_{S=1/2} (T,h)
    = \frac{1}{6}\xTTT
          - 0.00504 \xTTT ^{3}
         + O \xTTT^{5} ,
    \label{univ}
\end{equation}
from eqs.(\ref{chi1}) and (\ref{chi3}).

\subsection{Comparison with the Extended Scaling Function } \label{result-3}

In the previous subsection,
we get the scaling function (\ref{univ}) of
the $S=1/2$ quantum Heisenberg chain
by the Bethe Ansatz method.
In this subsection
we compare it
with the extended scaling function of eq.(\ref{arbS}).

In the case of $S$=1/2 in eq.(\ref{arbS}),
the extended scaling function is as follows:
\begin{eqnarray}
  F\left( \frac{ Jh}{4T^{2}} \right)
   &=& \frac{1}{6} \xTTT -\frac{11}{2160} \xTTT^3 + O\xTTT^5. \label{univclass}
\end{eqnarray}
The third order coefficient of the scaling function
given by the numerical calculation is
$-$0.00504 from eq.(\ref{univ}).
Since $(11/2160) \sim 0.00509,$
the extended scaling function of
eq.(\ref{univclass}) agrees
with the numerical results of
eq.(\ref{univ}) within the limits of an error.
\eqreset
\section{Summary and Discussion}\label{sum}
We propose the scaling relation between the magnetization,
the magnetic field and the temperature
for the spin $S$ ferromagnetic Heisenberg chain
in the analogy with the classical limit as follows:
\begin{equation}
 m_S(T,h)
 =  F\left( \frac{  S^2J h }{T^{2}} \right) ,\label{arbSS}
\end{equation}
where
\begin{equation}
F(x) =  \frac{2}{3} x - \frac{44}{135} x^{3}
                     + O (x^{5}) . \label{sumf}
\end{equation}

In the case of $S$=1/2,
we  calculate the Helmholtz free energy of the S=1/2 ferromagnetic Heisenberg
chain in a magnetic field by the
Bethe Ansatz method.
According to this calculation
the Helmholtz free energy
is expanded in respect of both $t^{0.5}$ and ${m}^{2}$ at
the low temperatures
and the low magnetization.
The coefficient of the first term $ t^{1.5} $,
which is independent of $m$, agrees
with the result given by integral equations\cite{TY}.
We get the $m$ dependence of the coefficients of
${t}^{2},{t}^{2.5}$ and ${t}^{3}$ up to the forth power $m$.
This results give us the third order nonlinear susceptibility.
Moreover we get not only $m$-$h$ curve but also the shape of the scaling
function.
Especially for small $h/(Jt^{2}),$
we get the coefficients of
the scaling function up to the third order.
We show that
this scaling function agrees with the proposed
scaling function.
According to our physical picture
we will be able to
obtain the higer order scaling function analytically
of the spin $S$ ferromagnetic Heisenberg chain
by calculating the higer order susceptibilities in the classical limit.

\vspace{1cm}
\begin{center}
{\bf ACKNOWLEDGEMENT}
\end{center}

The authors are grateful to Dr.Tohru Kawarabayashi
for stimulating discussions.
This work was supported in part by
Grants-in-Aid for Scientific Research on
Priority Areas, "Computational Physics as a New Frontier in Condensed
Matter Research" (Area No 217) and "Molecular Magnetism" (Area No 228),
from the Ministry of Education, Science and Culture, Japan.
\pagebreak
\eqreset
\eqreset
\appendix{The Nonlinear Susceptibility of the Classical Heisenberg Chain}
We consider the classical open Heisenberg chain
with classical spins ${\bspin}_{i}$
which are three dimensional unit vectors.
We choose a magnetic field {\boldmath $h$}
$(|{\mbox {\boldmath$h$}}|=h)$
as $z$-axis
in this three dimensional space.
The Hamiltonian of the classical Heisenberg chain is defined  as follows:
\begin{equation}
	{\cal H }^{\rm C}
  \equiv -J_{0}{\cal H }_{0}^{\rm C} - h {\cal H }_{1}^{\rm C},
\label{class_H}
\end{equation}
where
\begin{eqnarray}
{\cal H }_{0}^{\rm C} &=&  \sum_{i=1}^{N} {\bspin}_{i}\cdot
{\bspin}_{i+1},\label{h0} \\
{\cal H }_{1}^{\rm C} &=& \sum_{i=1}^{N} {r}_{i}^{z}.\label{h1}
\end{eqnarray}
The partition function $Z_{N}^{\rm C}$ is
\begin{equation}
    Z_{N}^{\rm C}
 \equiv \int \cdots \int \frac{d \Omega_{1}}{4 \pi} \cdots \frac{d
\Omega_{N}}{4 \pi} \exp(K {\cal H }_{0}^{\rm C}  +L {\cal H }_{1}^{\rm C}) \ ,
\label{classpart}
\end{equation}
where
\begin{equation}
K \equiv \frac{J_{0}}{T},L \equiv \frac{h}{T}, \label{coef}
\end{equation}
and $d\Omega_{i}$ is the element of solid angle for the vector $\bspin_{i}.$
Then the Gibbs free energy per site $f^{\rm C}$ is
\begin{equation}
f^{\rm C} = -\frac{T}{N} \ln Z_{N}^{\rm C} . \label{c_free}
\end{equation}

The linear susceptibility $\chi_{1}^{\rm C}$
and the nonlinear susceptibility $\chi_{3}^{\rm C}$
are defined as follows:
\begin{eqnarray}
\chi_{1}^{\rm C}
    &\equiv& \left. -\frac{{\partial}^{2}f^{\rm C}}{\partial h^{2}}
\right|_{h=0} ,\label{defchi1} \\
\chi_{3}^{\rm C}
    &\equiv& \left. -\frac{1}{3!}\frac{{\partial}^{4}f^{\rm C}}{\partial h^{4}}
\right|_{h=0} \  . \label{defchi3}
\end{eqnarray}
To evaluate these susceptibilities,
we must calculate
the 2-point and 4-point functions of this classical Heisenberg chain
without a magnetic field $({\mbox i.e.} h =0),$
which are defined by
\begin{eqnarray}
< r_{i}^{z} r_{j}^{z}> &\equiv&
     \frac{{\rm Tr} r_{i}^{z} r_{j}^{z} \exp(K{\cal H }_{0}^{\rm C} ) }{
            {\rm Tr}  \exp(K{\cal H }_{0}^{\rm C} ) }  , \label{2pointdef} \\
< r_{i}^{z} r_{j}^{z} r_{k}^{z} r_{l}^{z}> &\equiv&
     \frac{{\rm Tr} r_{i}^{z}
      r_{j}^{z} r_{k}^{z} r_{l}^{z}
          \exp(K{\cal H }_{0}^{\rm C} ) }{
            {\rm Tr}  \exp(K{\cal H }_{0}^{\rm C} ) }.   \label{4pointdef}
\end{eqnarray}
By using eqs.(\ref{classpart})-(\ref{4pointdef}),
the susceptibilities are derived as follows:
\begin{equation}
\chi_{1}^{\rm C} = \frac{1}{NT} A(K) , \label{achi1}
\end{equation}
\begin{equation}
\chi_{3}^{\rm C} = \frac{1}{3!NT^{3}} B(K)    ,  \label{nonlinearchi}
\end{equation}
where
\begin{eqnarray}
A (K) &\equiv&  \sum_{i=1}^{N} \sum_{j=1}^{N}
        < r_{i}^{z} r_{j}^{z}> ,   \label{ak} \\
B (K) &\equiv&
            \sum_{i=1}^{N} \sum_{j=1}^{N}
            \sum_{k=1}^{N} \sum_{l=1}^{N}
                   a(i,j,k,l),      \label{vk}
\end{eqnarray}
\begin{eqnarray}
  a(i,j,k,l) &\equiv&
          < r_{i}^{z} r_{j}^{z} r_{k}^{z} r_{l}^{z}>
            - < r_{i}^{z} r_{j}^{z}>
               < r_{k}^{z} r_{l}^{z}>      \nonumber \\
    & &   - < r_{i}^{z} r_{k}^{z}>
               < r_{j}^{z} r_{l}^{z}>
                 - < r_{i}^{z} r_{l}^{z}>
               < r_{j}^{z} r_{k}^{z}>  .
         \label{aa}
\end{eqnarray}
Fisher\cite{Fisher} gave the 2-point function as follows:
\begin{equation}
 <r_{i}^{z}r_{j}^{z}>
   = \frac{u^{-|i-j|}}{3} ,         \label{2point}
\end{equation}
where $u$ is Langevin function and $u \equiv \coth K - (1/K).$
Then we have $A(K)$ as follows:
\begin{eqnarray}
  A(K) &=& \left[
            2 \sum_{i < j}
          +  \sum_{i=j}
      \right] \frac{u^{j-i}}{3}                \nonumber \\
       &=& \frac{1}{3}
             \left(
                \frac{1+u}{1-u}
             \right) N +O(1)  .            \label{aaak}
\end{eqnarray}
As $T/J_0 \rightarrow 0$ (i.e. $K \rightarrow \infty $),
Langevin function behaves as follows:
\begin{equation}
u  \rightarrow 1- \frac{1}{K}. \label{Lang1}
\end{equation}
Therefore eqs.(\ref{achi1}),(\ref{aaak})
and (\ref{Lang1}) give us
the linear susceptibility\cite{Fisher}
at low temperatures as follows:
\begin{equation}
   \chi_{1}^{\rm C} \rightarrow \frac{2J_{0} }{3T^{2} } .\label{linearchi}
\end{equation}

Next we calculate the 4-point correlation function for the classical
Heisenberg chain without a magnetic field.
Let the polar angle of a spin ${\bspin}_{i}$ referred to the $z$-axis
be $\Theta_{i}.$
We define $\theta_{i+1}$  and $\phi_{i+1}$ as polar coordinates for
${\bspin}_{i+1}$ ,which are referred to ${\bspin}_{i}$ as the polar axis
with ${\bspin}_{1}$ defining
the reference plane for $\phi_{i+1}.$
We have
\begin{equation}
\cos \Theta_{i+1} = \cos \Theta_{i}
\cos \theta_{i+1} + \sin \Theta_{i}
\sin \theta_{i+1} \cos \phi_{i+1} . \label{angle}
\end{equation}
Here we show two recurrence relations to evaluate the 4-point correlation
function.
Let $i,$ $j,$ and $k$ be such integers as $1 \le i \le j \le k \le N.$
By using eq.(\ref{angle}) and $< \cos \phi_{i} > = 0,$
the first recurrence relation is
\begin{eqnarray}
<\cos \Theta_{i} \cdots \cos \Theta_{j} \cos \Theta_{k} > &=&
          <\cos \Theta_{i} \cdots \cos \Theta_{j} \cos \Theta_{k-1} >
< \cos \theta_{k} > \nonumber \\
      &\equiv&  <\cos \Theta_{i} \cdots \cos^{2}
\Theta_{j} >u^{k-j} \ . \label{form1}
\end{eqnarray}
We square eq.(\ref{angle})
and we obtain the second recurrence relation
as follows:
\begin{eqnarray}
<\cos^{2} \Theta_{i} \cos^{2} \Theta_{j} >
  &=&  <\cos^{2} \Theta_{i} \cos^{2} \Theta_{j-1}> v + \frac{u}{3K}  \nonumber
\\
  &=& \frac{1}{9} \left(\frac{4v^{j-i}}{5} + 1\right)    ,      \label{form2}
\end{eqnarray}
where $v \equiv 1-(3u/K)$.
We get the 4-point function by eqs.(\ref{4pointdef}), (\ref{form1}) and
(\ref{form2}) as follows:
\begin{eqnarray}
<r_{i}^{z}r_{j}^{z}r_{k}^{z}r_{l}^{z}>
   &=& <\cos \Theta_{i} \cos \Theta_{j}
\cos \Theta_{k} \cos \Theta_{l} > \nonumber \\
   &=& u^{j-i}<  \cos^{2} \Theta_{j}\cos^{2} \Theta_{k}>u^{l-k} \nonumber \\
   &=& \frac{1}{9} u^{j-i} \left(\frac{4v^{k-j}}{5} + 1\right) u^{l-k} .
        \label{4point}
\end{eqnarray}
By using eqs.(\ref{aa}), (\ref{2point}) and (\ref{4point}) we have
\begin{equation}
a(i,j,k,l)= \frac{1}{9} u^{j-i}\left\{ \frac{4}{5}v^{k-j}-2u^{2(k-j)} \right\}
u^{l-k}
            , \label{aaa}
\end{equation}
where $i \le j \le k \le l.$


We consider $B(K)$
to calculate $\chi_{3}^{\rm C}.$
By using eqs.(\ref{vk}) and (\ref{aaa}), we sum up
$a(i,j,k,l)$ for all cases of $\{i,j,k,l\}$ as follows:
\begin{eqnarray}
& & B (K)  \nonumber \\
  &=& \left[
            24 \sum_{i \le j<k<l}
          + 12 \sum_{i<j=k<l}
          + 6  \sum_{i=j<k=l}
          + 8  \sum_{i=j=k<l}
          + \sum_{i=j=k=l}
      \right] a(i,j,k,l)                  \nonumber \\
  & & =  \left\{
          \frac{8}{15} \frac{v}{1-v} {\left( \frac{1+u}{1-u} \right) }^{2}
        - \frac{2(5u^{2}+4u+1)}{15( 1-u  )^{2} } \left( \frac{1+u}{1-u} \right)
      \right\} N
      + O(1)   .       \nonumber \\
  & & \label{DELTA}
\end{eqnarray}
As $T/J_0 \rightarrow 0$ (i.e. $K \rightarrow \infty $),
$v$ behaves as follows
\begin{eqnarray}
v&\rightarrow& 1- \frac{3}{K}.\label{Lang2}
\end{eqnarray}
Then the asymptotic behavior of $B(K)$ is
\begin{equation}
 B (K)  \rightarrow  - \frac{88}{45} K^{3} N         .\label{DELTA2}
\end{equation}
Using eqs.(\ref{coef}), (\ref{nonlinearchi}) and (\ref{DELTA2}),
we get the nonlinear susceptibility  for large $K$ (i.e. low $T$)
as follows:
\begin{equation}
\chi_{3}^{\rm C}
   \rightarrow - \frac{44J_{0}^{3}}{135T^{6}} . \label{chiinfty}
\end{equation}
\pagebreak

\pagebreak
{\bf Figure Captions}

\begin{enumerate}
 \item
The $ t^{0.5} $ dependence of $ g (t,m)/(J{t}^{1.5})$ for $ m = 0,$ 0.28,
0.632.
As $t \rightarrow 0 $, all curves approach to the same value $-1.04218$
(filled square). The curves are a guide for eyes.
  \label{free_ene}
 \item
When $m=0$ in eq.(\ref{g}), $a_{0}(t) = g(t,0)/(Jt^{1.5}).$
We plot $a_{0}(t)$ versus $t^{0.5}.$
A solid curve is given by the least-squares method,
which is $-1.04218 + 1.00 t^{0.5} - 0.94t + 0.9 t^{1.5}. $
  \label{keisu1}
 \item
When we expand $h/(Jmt^{2}) $ with
$m^{2}$ for fixed $t$,
we can get expansion coefficients.
We plot the first order coefficient $2a_{1}(t)$ versus
$t^{0.5}$ by the same method as in Fig.\ref{keisu1}.
A solid curve is given by the least-squares method,
which is $1/\{ (1/6)   + 0.5826 t^{0.5} + 0.678 t + 0.04 t^{1.5} \}. $
  \label{keisu2}
 \item
We plot the second order coefficient $4a_{2}(t)$ versus $t^{0.5}$
by the same method as Fig.\ref{keisu1}.
A solid curve is given by the least-squares method,
which is $1/\{ 0.153  + 0.83 t^{0.5} + 2.4 t \}. $
  \label{keisu3}
\item
We plot $\ln \{h/(Jt^{2} )\}$ versus $t^{0.5}$ for $m =
 0.01,$ 0.04, 0.06, 0.1, 0.14, 0.28, 0.5, 0.63.
A quadratic function is used as our fitting function.
Results are expressed by solid lines.
Their $y$-intercepts (empty circles) are
limitation values of $\ln \{h/(Jt^{2})\}$ as
 $t \rightarrow 0.$
\label{m_hcurve}
\item
We plot $m$ versus $\ln \{h/(Jt^{2} )\}$ by filled circles for $t =$
0.2, 0.1, 0.05, 0.025, 0.005.
We draw fine curves as a guide for eyes.
When $t \rightarrow 0,$
we plot the limitation values (empty squares) given in Fig.\ref{m_hcurve}.
A bold curve is drawn as a guide for eyes.
As we increase a number of the limitation values,
the bold curve is getting to the scaling function.
\label{m_hcurve2}
\end{enumerate}
\end{document}